\documentclass[aps,prb,showpacs,twocolumn,longbibliography,10pt]{revtex4-1}
\pdfoutput=1
\usepackage{graphicx}
\usepackage{amsmath}
\usepackage{amsbsy}
\usepackage{hyperref}
\usepackage{dcolumn}
\usepackage{bm}
\usepackage{color}
\newcommand{\mbf}{\mathbf} 
\renewcommand{\k}{{\mbf k}}

\newcommand{\q}{{\mbf q}}
\def\colorb{}
\def\colorr{}

\begin{document}

\title{Itinerant approach to magnetic neutron scattering of FeSe: effect of orbital selectivity}
\author{Andreas Kreisel$^1$, Brian M. Andersen$^2$, and P. J. Hirschfeld$^3$}
\affiliation{
$^1$Institut f\" ur Theoretische Physik, Universit\" at Leipzig, D-04103 Leipzig, Germany\\
$^2$Niels Bohr Institute, University of Copenhagen, Lyngbyvej 2, DK-2100 Copenhagen, Denmark\\
$^3$Department of Physics, University of Florida, Gainesville, Florida 32611, USA}

\date{January 7, 2019}
\begin{abstract}

Recent STM experiments and theoretical considerations have highlighted the role of interaction-driven orbital selectivity in FeSe, and its role in generating the extremely anisotropic superconducting gap structure in this material. 
We study the magnetic excitation spectrum resulting from the coherent  quasiparticles within the same renormalized random phase approximation approach used to explain the STM experiments, and show that it agrees well with the low-energy momentum and energy dependent response measured by inelastic neutron scattering experiments.  We find a correlation-induced suppression of $(\pi,\pi)$ scattering due to a small quasiparticle weight of states of  $d_{xy}$ character.  We compare predictions for twinned and untwinned crystals,  and  predict in particular a strongly $(\pi,0)$-dominated response at low energies in untwinned systems, in contrast to previous itinerant theories.

\end{abstract}


\maketitle
\section{Introduction}
Among the various iron-based superconductors (FeSC), the compound FeSe has been in the focal point of research recently.  
Although the bulk material has a rather low critical superconducting transition temperature $T_c$   of approximately 8\;\text{K}~\cite{Hsu2008}, there are many ways to enhance $T_c$.
Intercalation with various other atoms and molecules yields $T_c$ near 40\;\text{K}~\cite{Burrard-Lucas2013,Ying2011II,Krzton-Maziopa2016}, and a similar enhancement is observed upon application of pressure~\cite{Mizuguchi2008,Medvedev2009,Margadonna2009,Garbarino2009,Masaki2009,Okabe2010}. Remarkably, monolayer films of FeSe on SrTiO$_3$ reach critical temperatures around 70-100\;\text{K}~\cite{Wang12,Sadovskii2016,Wang2017,Huang2017,Ge2015}.
In addition, FeSe displays an interesting interplay between superconductivity, magnetism and nematic order, exhibiting a structural transition from tetragonal to orthorhombic crystal structure at $T_s\approx 90\;\text{K}$\cite{Hsu2008,Margadonna2008,McQueen2009}, without ordering magnetically\cite{McQueen2009,Bendele2010}. FeSe is a unique system to study the nematic phase, which in other FeSC only exists within a rather narrow temperature range, but the origin of the nematicity remains unclear at present~\cite{BoehmerKreisel_review,Pustovit2016}.

It is generally believed that an understanding of the properties  of bulk FeSe may lead to a clarification of the various routes to $T_c$ enhancement and possibly enable further improvement.   Recent advances include the determination of the structure of the magnetic phase that forms  upon application of pressure\cite{Terashima2015,Kothapalli2016,Wang2016}, the mapping of the phase diagram under both pressure and S substitution\cite{Matsuura2017,Coldea2016}, and measurements probing the low-temperature quasiparticle excitations in both the normal and superconducting states\cite{Kasahara2014,Teknowijoyo2016,Li2016,Bourgeois-Hope2016,Sprau2017,Kostin2018}. 

For clues to the reasons for suppression of magnetic order, as well as the origin of unconventional superconductivity in FeSe and related materials, an important measurement is  inelastic neutron scattering (INS), which probes elementary spin excitations via the imaginary part of the dynamical spin susceptibility $\chi''({\bf q},\omega)$ \cite{Rahn2015,Wang2015,Wang2015GS}. These results showed the presence of strong stripe-like (${\bf q}=(\pi,0))$ spin fluctuations at low energies, as well as a superconducting state resonance at the same wave vector, similar to FeSC compounds that do order magnetically.  However, in addition they found unusually strong N\'eel-type (${\bf q}=(\pi,\pi)$)  fluctuations whose amplitude decreased relative to the ($\pi,0$) fluctuations as temperature decreased below the structural transition temperature $T_s$. Ref. \onlinecite{Wang2015GS} further showed the existence of a significant spin gap of 30-40 meV for $(\pi,\pi)$ excitations at low temperatures.

From a theoretical standpoint, the lack of magnetic order had been discussed even before these measurements appeared,  in terms of frustration among various possible magnetically ordered states in extended quantum spin models 
\cite{Glasbrenner2015,WangLee2015NatPhys_FeSe-Para-Nematicity,Liu_16,Yu2015} including Heisenberg and biquadratic spin exchanges.  While the fluctuating moment $\langle m^2 \rangle$ of the Fe ion is indeed close to $\sim 5\mu_B^2$, the system is itinerant; in principle a complete description of the magnetic properties should therefore explain how the observed magnetic excitations, along with their evolution in frequency and temperature, arise from the original Fe $d$-electrons.  Indeed, the authors of Refs. \onlinecite{Glasbrenner2015,Liu_16,Busemeyer_16} proposed that conventional density functional theory (DFT) calculations suggested a competition between nearly degenerate magnetic ground states, and showed that DFT estimates of  magnetic exchanges in FeSe placed the system close to the boundary between several ordered magnetic states.  To explain the lack of long range magnetic order in FeSe other authors have focused on the role of vertex corrections\cite{Yamakawa_PRX}, or proposed that correlations give rise to a non-ideal orbital weight distribution at the Fermi surface\cite{Fanfarillo_mismatch2018}.  Finally, there have been some further suggestions of hidden quadrupolar magnetic order\cite{Yu2015,Wang_Z_16}, but to our knowledge there is no evidence that definitively supports this proposal.

The current authors  performed a random-phase-approximation (RPA) study of the magnetic susceptibility of the paramagnetic system\cite{Kreisel15}, employing a tight-binding band structure for the $d$-bands near the FeSe Fermi surface with coefficients chosen to fit angle-resolved photoemission (ARPES) and quantum oscillation (QO) experiments, as well as an orbital ordering term in the Hamiltonian with mean-field-like temperature dependence assumed.  These calculations provided a good description of many properties of the observed spin excitations, including the existence of both stripe-like and N\'eel fluctuations \footnote{A small $(\pi,\pi)$ scattering has been achieved by a renormalization of a hopping matrix element for the $d_{x^2-y^2}$ orbital that affects the electronic structure mostly at energies away from the Fermi level, thus leaving the Fermi surface unaffected.}, as well as the qualitative features of their $T$- and $\omega$-dependences, suggesting that the low-energy spin excitations were indeed itinerant, paramagnon-like entities.  

Certain features of the experiment were not matched perfectly in Ref. \onlinecite{Kreisel15}, however: the low-energy, high temperature excitations near $(\pi,\pi)$ were found to be somewhat incommensurate, in contrast to the results of  Ref. \onlinecite{Wang2015GS}.  In addition, the low temperature, low energy $(\pi,\pi)$ fluctuations were gapped in experiment, whereas they had significant weight down to the lowest energies in the calculation.  Finally, the prediction for the structure of the superconducting gap given in Ref. \onlinecite{Kreisel15}, based on a calculation of the spin fluctuation pairing interaction within the same framework,  was not borne out by a recent high-resolution quasiparticle interference experiment by Sprau {\it et al}.\cite{Sprau2017}

These aspects were criticized by She {\it et al.}\cite{She2017arXiv}, who proposed a frustrated spin-1 picture consistent with the ideas of Refs. \onlinecite{Glasbrenner2015} and \onlinecite{WangLee2015NatPhys_FeSe-Para-Nematicity}.  Within a spin-fermion model, the authors introduced a Schwinger boson representation for the localized spins, and ignored the Bose condensate in order to describe the disordered quantum nematic spin liquid. They found a spin excitation spectrum with a commensurate $(\pi,\pi)$ gap, {\colorr made a sharp prediction for a strong anisotropy between $(\pi,0)$ and $(0,\pi)$}, and pointed out that if the Hund's coupling were orbitally dependent, a particular choice of these couplings could reproduce the FeSe superconducting gap as found in   Sprau {\it et al}\cite{Sprau2017}.    Recently, calculations employing exact diagonalization of $J_1-J_2-J_3-K$ Heisenberg model clusters also claimed qualitative agreement with the neutron data when the Hamiltonian parameters were chosen near a frustration point, suggesting that a localized spin picture was sufficient \cite{Baum2017}, but this work was not able to calculate spin excitations at low energies with high resolution. {\colorr Finally, it was noted that a system with  antiferroquadrupolar order also presents a strong distinction between the susceptibility at $(\pi,0)$ and $(0,\pi)$ in the untwinned case\cite{Yu2015,Lai2017}.}

The traditional RPA approach to calculating spin excitations with perfectly well-defined electronic dispersions may indeed be expected to encounter difficulties in systems with strong correlations, particularly for the iron chalcogenides\cite{Yin2011}.
The  calculations presented in Ref. \onlinecite{Kreisel15} incorporated some electronic correlation effects indirectly, for example via renormalization of the bare DFT-derived tight-binding band structure  to fit ARPES ~\cite{Yi2015} and quantum oscillations\cite{Terashima2014,Watson2015,Audouard2015EPL_Hc2}.  
This approach amounts to including the self-energy partially, i.e. shifting the pole of the Green's function according to $\mathop{\text{Re}}\Sigma(\k,\omega)$
near the Fermi surface. However, Ref. \onlinecite{Kreisel15}, neglected the suppression of quasiparticle weights $Z$ that can be a significant effect in correlated Fermi liquids {\colorb in general. In the context of Fe-based superconductors, a reduction of quasiparticle weights has been discussed in the context of orbital-selective Mottness\cite{Georges_rev_2013,deMedici2014,Biermann_review,Yi2017,Guterding2017,Medici_review}. However, in the following we do not make a direct connection to the localization in such a correlated state when using an itinerant approach.}  The importance of 
{\colorb the quasiparticle}weights for low-energy properties of the FeSe system was emphasized in the analysis of recent QPI data~\cite{Sprau2017,Kostin2018}. Based on the QPI, a mechanism for the formation of Cooper pairs was proposed which takes into account the reduction due to correlations of the quasiparticle weights of the electronic quasiparticles in FeSe~\cite{Kreisel2017}. As discussed initially within dynamical mean field theory (DMFT)\cite{Yin2011,Biermann_review} and confirmed qualitatively by ARPES\cite{Yi2017}, the FeSC are systems with moderately correlated electronic structure that exhibit stronger renormalizations of certain orbital states\cite{Yin2011,Guterding2017}; in particular, the $d_{xy}$ orbital quasiparticle weight is generically strongly suppressed as seen by a large mass renormalization of the $d_{xy}$-dominated band. Furthermore, in the nematic FeSe system, the renormalizations of the $d_{xz}$ and the $d_{yz}$ orbitals have been proposed to be significantly different\cite{Sprau2017,Kreisel2017,Hu2018,Fanfarillo2017}.  Indeed, ARPES orbital polarization analysis on untwinned crystals is consistent with this hypothesis\cite{Suzuki2015,Watson2017,Coldea_2017_review}, to the extent that the electron pocket of $d_{xz}$ character has proven very difficult to observe~\cite{Watson2017}.

In this work, we revisit the calculation of Ref. \onlinecite{Kreisel15}, and examine the physical magnetic susceptibility {\colorb in presence of interactions} measured by neutrons, incorporating now the orbital selective quasiparticle weights found in the recent studies of Refs. \onlinecite{Sprau2017,Kreisel2017,Kostin2018}.  We find that the earlier low-energy discrepancies with the INS experiments are repaired without further fine tuning, and propose therefore that an itinerant approach, properly accounting for Fermi liquid renormalizations, is indeed the most complete theory of the FeSe nematic state, reproducing the features that seemed to be restricted to localized spin models. In addition, we perform calculations appropriate both for twinned samples and for untwinned ones.  We predict that the effect of orbital selective quasiparticle weight is to strongly suppress the spin excitation intensity at ${\bf q}=(0,\pi)$ compared to $(\pi,0)$. Finally, we explore the consequences of the quasiparticle weights for other low-temperature properties such as the neutron spin resonance in the superconducting state.

\section{Model}

\begin{figure}[tb]
\includegraphics[width=\linewidth]{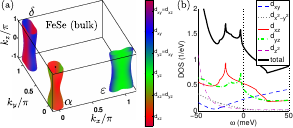}
\caption{(a) Fermi surface of our  model assuming coherent quasiparticles ($Z_\ell=1$) for FeSe showing orbital content of the bands in false color;  (b) the corresponding density of states at low energies.}
\label{fig_fermi}
\end{figure}

In this work, we use a tight-binding parametrization proposed for FeSe earlier\cite{Eschrig09,Sprau2017,Kreisel2017}

\begin{equation}
H=\sum_{\mathbf{k}\sigma \ell \ell'} t^{\ell \ell'}_{\mathbf{k}} c_{\ell\sigma}^\dagger(\k) c_{\ell'\sigma}(\k),
 \label{eq_tb}
\end{equation}

where $c_{\ell\sigma}^\dagger(\k)$ is the Fourier amplitude of an operator $c_{i\ell\sigma}^\dagger$ that creates an electron in  Wannier orbital $\ell$ with spin $\sigma \in \{\uparrow, \downarrow\}$ and $t^{\ell \ell'}_{\mathbf{k}}$ is the Fourier transform of the hoppings.   Specifically, we use the hoppings given in Ref. \cite{Sprau2017}.  Here, $\ell$ is an orbital index with $\ell\in(1,\ldots,5)$ corresponding to the Fe $3d$ orbitals $(d_{xy},d_{x^2-y^2},d_{xz},d_{yz},d_{3z^2-r^2})$.

Note that in the kinetic energy of this
model we have also included  spin-orbit coupling of  type $S^zL^z$, which gives rise to imaginary hopping elements onsite between different orbitals. Details of the corresponding elements have been discussed in Ref.~\onlinecite{Kreisel13} and the consequences of such a term, namely the splitting of the two hole-like bands along the line $\Gamma$-$Z$ in the Brillouin zone have been discussed earlier\cite{Sprau2017,Kreisel2017} and are in line with experimental findings\cite{BorisenkoSO}. At low temperatures, FeSe is found to be nematic and the corresponding splittings of the eigenenergies have been modeled by the introduction of an onsite potential in the $d_{xz}$ and $d_{yz}$ orbitals (site order $\Delta_s$) and a bond order term $\Delta_b$ which changes the hoppings of those orbitals to the nearest neighbor positions (bond order). 
{The orbital order and the associated splittings of the band structure can be  captured theoretically at a semi-quantitative level\cite{Jiang_16,Wu_2016arXiv160302055W,Scherer_FeSe}.}
In momentum space at low temperatures, this term then reads

\begin{align}
 H_{OO}&=\Delta_b \sum_{\mathbf k}(\cos k_x -\cos k_y) (n_{xz}({\mathbf k})+n_{yz}({\mathbf k}))\notag\\
 & +\Delta_s\sum_{\mathbf k}(n_{xz}({\mathbf k})-n_{yz}({\mathbf k}))\,,
  \label{eq_oo}
\end{align}
where $n_\ell({\mathbf k})=\sum_\sigma c_{\ell\sigma}^\dagger(\k) c_{\ell\sigma}(\k)$. 
The Fermi surface and the density of states of this model are illustrated in Fig. \ref{fig_fermi}.
Recently, Ref. \onlinecite{Yu2018} suggested that an additional term of type $\Delta_s'(\cos k_x+\cos k_y)(n_{xz}(\k) - n_{yz}(\k))$, could be important, but we do not consider such effects here.
A unitary transformation with the matrix elements $a_\mu^\ell(\k)$ diagonalizes the Bloch Hamiltonian such that it becomes
\mbox{$H=\sum_{\mathbf{k}\sigma \mu}\tilde E_\mu(\k) c_{\mu\sigma}^\dagger(\k)c_{\mu\sigma}(\k)$}, with eigenenergies $\tilde E_\mu(\k)$ that match the maxima of the spectral function as deduced experimentally\cite{Terashima2014,Audouard2015EPL_Hc2,Watson2015,Watson2016,Sprau2017,Kostin2018} and $c_{\mu\sigma}^\dagger(\k)$ creating an electron in Bloch state $\mu,\k$.
Together with the experimental observation of the band structure $\tilde E_\mu(\k)$,  properties of the spectral function $\tilde A(\k,\omega)$ can be deduced. We parametrize the Green's function in orbital space as
\begin{align}\label{eq_GF}
\tilde G_{\ell\ell'} (\k,\omega_n)&=\sqrt{Z_\ell  Z_{\ell'}} \;\sum_\mu \frac{a_\mu^\ell(\k) a_\mu^{\ell'*}(\k)}{i \omega_n - \tilde E_\mu(\k)}\notag\\
&=\sqrt{Z_\ell  Z_{\ell'}} \;\sum_\mu a_\mu^\ell(\k) a_\mu^{\ell'*}(\k)\tilde G^\mu(\k,\omega_n),
\end{align}
with quasiparticle weights $Z_\ell$ in orbital $\ell$ and we have introduced the (coherent) Green's function in band space as $\tilde G^\mu(\k,\omega_n)=[{i \omega_n - \tilde E_\mu(\k)}]^{-1}$. Note the Green's function $\tilde G$, which we refer to as renormalized or incoherent because the quasiparticle weight factors $Z<1$ are included, does {\it not} include the actual incoherent spectral weight, {\colorb such that  {\colorr the resulting total electron density is not calculated consistently}. We note further that the self-energy in general also exhibits a nonzero imaginary part which is expected to grow quadratically with the energy measured from the Fermi level.  Our parametrization} is therefore simply a phenomenological ansatz that agrees with the low energy properties of the spectral function $\tilde A(\k,\omega)=-1/\pi\mathop{\text{Im}}\sum_\ell \tilde G_{\ell\ell} (\k,\omega_n)$ of the real material\cite{Kostin2018,Rhodes2018}, {\colorb but contains effects of correlations and properties of the electronic nematicity beyond results from DFT based methods in this system.}

We include local interactions via the standard Hubbard-Hund Hamiltonian  
\begin{eqnarray}
	H &=& {U}\sum_{i,\ell}n_{i\ell\uparrow}n_{i\ell\downarrow}+{U}'\sum_{i,\ell'<\ell}n_{i\ell}n_{i\ell'}
	\nonumber\\
	& + & {J}\sum_{i,\ell'<\ell}\sum_{\sigma,\sigma'}c_{i\ell\sigma}^{\dagger}c_{i\ell'\sigma'}^{\dagger}c_{i\ell\sigma'}c_{i\ell'\sigma}\nonumber\\
	& + & {J}'\sum_{i,\ell'\neq\ell}c_{i\ell\uparrow}^{\dagger}c_{i\ell\downarrow}^{\dagger}c_{i\ell'\downarrow}c_{i\ell'\uparrow} , \label{H_int}
\end{eqnarray}
where the parameters ${U}$, ${U}'$, ${J}$, ${J}'$ are given in the notation of Kuroki \textit{et al.} \cite{Kuroki2008} Imposing spin-rotational invariance, i.e. $U'=U-2J$, $J=J'$, there are only two parameters $U$ and $J/U$ left to specify the interactions. 

Note that these interactions will lead to a self-energy $\Sigma_{\ell\ell'} (\k,\omega_n)$ which induces both a shift of the eigenenergies and a reduced quasiparticle weight at the Fermi level as we model it with our ansatz. The $\k$-dependent shift of the bands is effectively included already in our renormalized band structure $\tilde E_\mu({\bf k})$ since it is fit to experiment, but the $Z$-factors must be included explicitly\cite{Kreisel2017} via our ansatz (\ref{eq_GF}).    A microscopic calculation of the self-energy is in progress, but is beyond the scope of this paper.

Within the current ansatz, two-particle properties are renormalized in similar ways; for example,  the orbital susceptibility in the normal state is

\begin{align}
	\tilde\chi_{\ell_1 \ell_2 \ell_3 \ell_4}^0 (q) & = - \sum_{k,\mu,\nu }\tilde M_{\ell_1 \ell_2 \ell_3 \ell_4}^{\mu\nu} (\k,\q) \tilde G^{\mu} (k+q) \tilde G^{\nu} (k),   \label{eqn_supersuscept}
\end{align}
where we have adopted the shorthand $k\equiv (\k,\omega_n)$ and defined the abbreviation
\begin{eqnarray}
	\tilde M_{\ell_1 \ell_2 \ell_3 \ell_4}^{\mu\nu} (\k,\q)& =&\sqrt{Z_{\ell_1}Z_{\ell_2}Z_{\ell_3}Z_{\ell_4}}   \\&&\times a_\nu^{\ell_4} (\k) a_\nu^{\ell_2,*} (\k) a_\mu^{\ell_1} (\k+\q) a_\mu^{\ell_3,*} (\k+\q).\nonumber
\end{eqnarray}

The internal frequency summation can be  performed analytically and we calculate $\tilde\chi_{\ell_1 \ell_2 \ell_3 \ell_4}^0$ by integrating over the full Brillouin zone.
Actually, it is easy to see, that one can calculate the susceptibility $\chi_{\ell_1 \ell_2 \ell_3 \ell_4}^0$ using a fully coherent Green's function (without the quasiparticle weight prefactors in Eq. (\ref{eq_GF})) such that
the two quantities are just related by the equation
\begin{equation}
 \tilde \chi_{\ell_1 \ell_2 \ell_3 \ell_4}^0 (\q)=\sqrt{Z_{\ell_1}Z_{\ell_2}Z_{\ell_3}Z_{\ell_4}}\;\chi_{\ell_1 \ell_2 \ell_3 \ell_4}^0 (\q),
 \label{eq_susc}
\end{equation}
i.e. the quasiparticle weights just enter as prefactors {\colorb and suppress certain orbital channels, thereby altering the momentum structure of this quantity straightforwardly. This leads to a relative suppression of the susceptibility in the $d_{xy}$ channel compared to the $d_{yz}$ channel with the quasiparticle weights employed\cite{Kreisel2017} as also found in recent DMFT investigations on FeSe\cite{Ishizuka18}.}
When one transforms to the band basis, these renormalizations of course acquire an additional momentum dependence.

Turning now to the consequences of the low-energy parametrization, it has been discussed previously\cite{Kreisel2017} that the bare  static  susceptibility in the fully coherent case has overall similar magnitude in the $\q=(\pi,\pi)$ versus $\q=(\pi,0)$ channels but with orbitally distinct momentum structure. This can be partially understood by the different nesting conditions at low energies for the different orbital components: The $d_{xy}$ component,  present predominantly on the electron pockets, has a maximum at $\q=(\pi,\pi)$ due to scattering with this momentum transfer, while the components for $d_{yz}$ ($d_{xz}$) have maxima at $\q=(\pi,0)$ ($\q=(0,\pi)$) from similar dominant scattering between the hole-like band and the electron-like band at the X point (Y point). Note that the influence of the orbital order, as presented in Eq.(\ref{eq_oo}), on the eigenenergies is rather small compared to energy scales of the electronic structure, i.e. the bandwidth $W\gg\Delta_s=9.6\text{meV}$, $W\gg |\Delta_b|=|-8.9\text{meV}|$, thus the resulting differences in the susceptibility  between the X and Y points from a conventional coherent calculation are weak as well\cite{Kreisel2017}.

Employing now the quasiparticle weights as deduced from a calculation of the superconducting order parameter\cite{Sprau2017} and in agreement with considerations from observed anisotropies in the scattering amplitudes in the normal state of FeSe\cite{Kostin2018}, i.e. fixing $\{\sqrt{Z_l}\}=[0.2715,0.9717,0.4048,0.9236,0.5916]$  (``$Z_\ell$ set I''  for orbitals $xy$, $x^2-y^2$, $xz$, $yz$, $z^2$ ) and using Eq. (\ref{eq_susc}), it is apparent that certain components of the bare susceptibility will be suppressed, with important consequences for the momentum structure of the physical spin susceptibility changing the dominant low-energy weight from $(\pi,\pi)$ to $(\pi,0)$. 

 We stress that the actual numbers for $Z_l$ written above are subject to large uncertainties, and will depend also to some extent on the initial bare band. However, it appears clear that the $d_{xy}$-dominated ($\pi,\pi$) scattering needs to be suppressed, and that additional (compared to a nematic but fully coherent scenario) anisotropy in the quasi-particle weights between $d_{xz}$ and $d_{yz}$ is also necessary for explaining both the normal state QPI and the superconducting gap structure within a modified spin-fluctuation theory.

\begin{figure}[tb]
\includegraphics[width=\columnwidth]{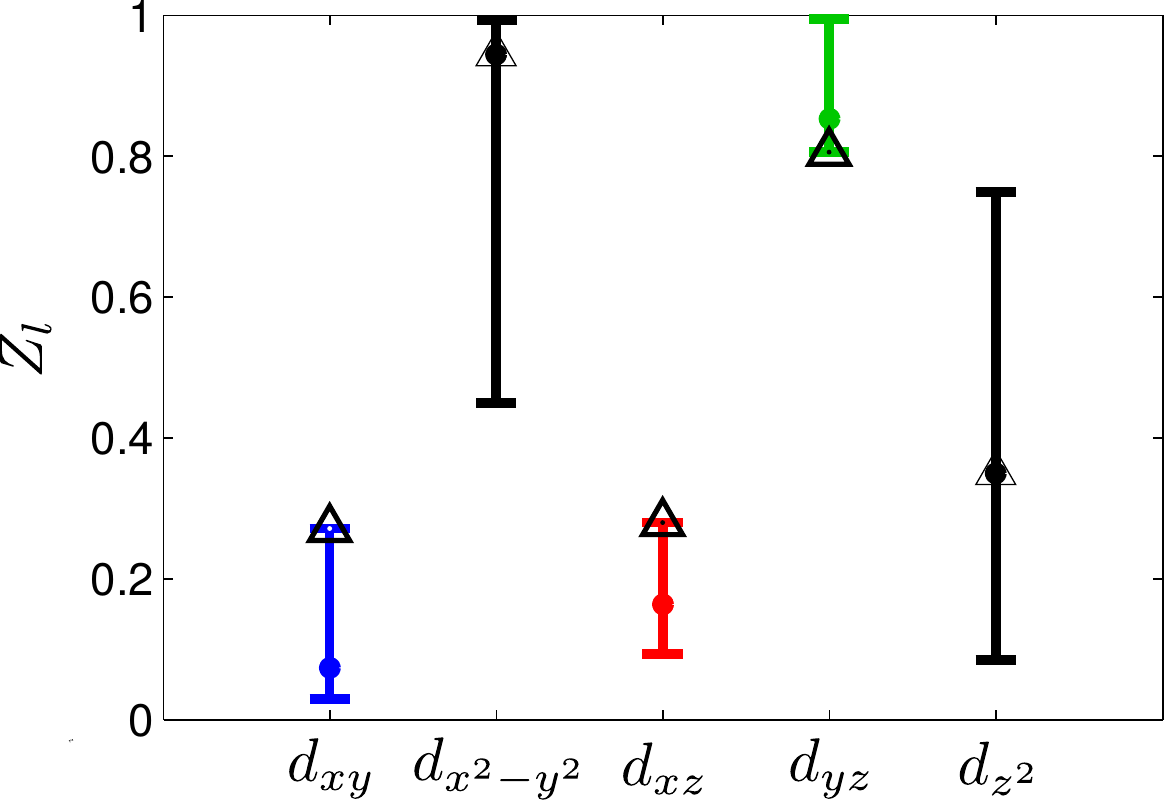}
\caption{Values of the quasiparticle weights $Z_l$ set I as obtained from the calculation of the superconducting order parameter (dots). The corresponding error bars have been obtained by changing one quasiparticle weight such that the reduced $\chi^2$ in the fit to the QPI-determined gap function in Ref. \onlinecite{Sprau2017} increased by one. An alternative set of quasiparticle weights ($Z_\ell$ set II) within this range of the error bars has been used for calculations as well  (black triangles)}.
\label{Z_bars}
\end{figure}

 In this paper, we focus primarily on the set of $Z_\ell$'s given above that provided the best fit to the superconducting gap in Ref. \onlinecite{Sprau2017}, as a consistency check.  However it is clear that a range of such choices is consistent with the data in Refs. \onlinecite{Sprau2017} and \onlinecite{Kostin2018}.  To give a rough idea of what $Z$'s are possible for FeSe within the current framework, we have compared the calculated superconducting gap with the experimental error bars from Ref. \onlinecite{Sprau2017}, varying each $Z_\ell$
independently.  Fig. \ref{Z_bars} shows the allowed values according to this procedure.  It is clear that somewhat larger error bars would be allowed if correlated variations of the $Z$'s were taken.  For this work, due to computational restrictions, we consider only the ``canonical'' set given above from Ref. \onlinecite{Sprau2017} and, where indicated, a second set  $\{\sqrt{Z_l}\}=[0.5215,0.9717,0.5298,0.8986,0.5916]$  (``$Z_\ell$ set II'') with  somewhat less suppression of $Z_{xy}$  and reduced differentiation between $Z_{xz}$ and $Z_{yz}$.    This set, marked with  black  triangles in Fig. \ref{Z_bars}, is  still consistent with the determination of the superconducting gap in Ref. \onlinecite{Sprau2017}, as shown.  Note that the values of $Z_{x^2-y^2}$ and $Z_{z^2}$ are essentially irrelevant since the corresponding $d$-states have no weight at the Fermi level in FeSe.

Interactions (Eq. (\ref{H_int})) can be incorporated into two-particle properties via the random-phase approximation (RPA) by summing a subset of diagrams to evaluate the full response function (see, e.g. Ref. \onlinecite{Graser2009}).
The spin-fluctuation part of the RPA susceptibility, $\tilde\chi_1^{\rm RPA}$, is given as

\begin{align}
\label{eqn:RPA}
 \tilde\chi_{1\,\ell_1\ell_2\ell_3\ell_4}^{\rm RPA} (\q,\omega) &= \left\{ \tilde\chi^0 (\q,\omega) \left[1 -\bar U^s \tilde\chi^0 (\q,\omega) \right]^{-1} \right\}_{\ell_1\ell_2\ell_3\ell_4}.
\end{align}
The interaction matrix $\bar U^s$ in orbital space is composed of linear combinations of $U,U',J,J'$ and its form is given, e.g., in Ref.~\onlinecite{a_kemper_10}.
The total physical spin susceptibility  is then given by the sum
\begin{equation}
	\label{eqn_chisum} \chi_\text{RPA} (\q,\omega) = \frac 12 \sum_{\ell \ell^\prime} \tilde\chi_{1\;\ell \ell \ell^\prime\ell^\prime}^{\rm RPA} (\q,\omega)=\sum_{\ell}\chi_{\ell} (\q,\omega)\,.
 \end{equation}
 Similarly, the spin fluctuation pairing interaction is obtained from the usual expressions\cite{Graser2009} with the substitution $\chi^0\rightarrow \tilde \chi_0$\cite{Kreisel2017}.

\section{Results}

\begin{figure}[tb]
\includegraphics[width=\linewidth]{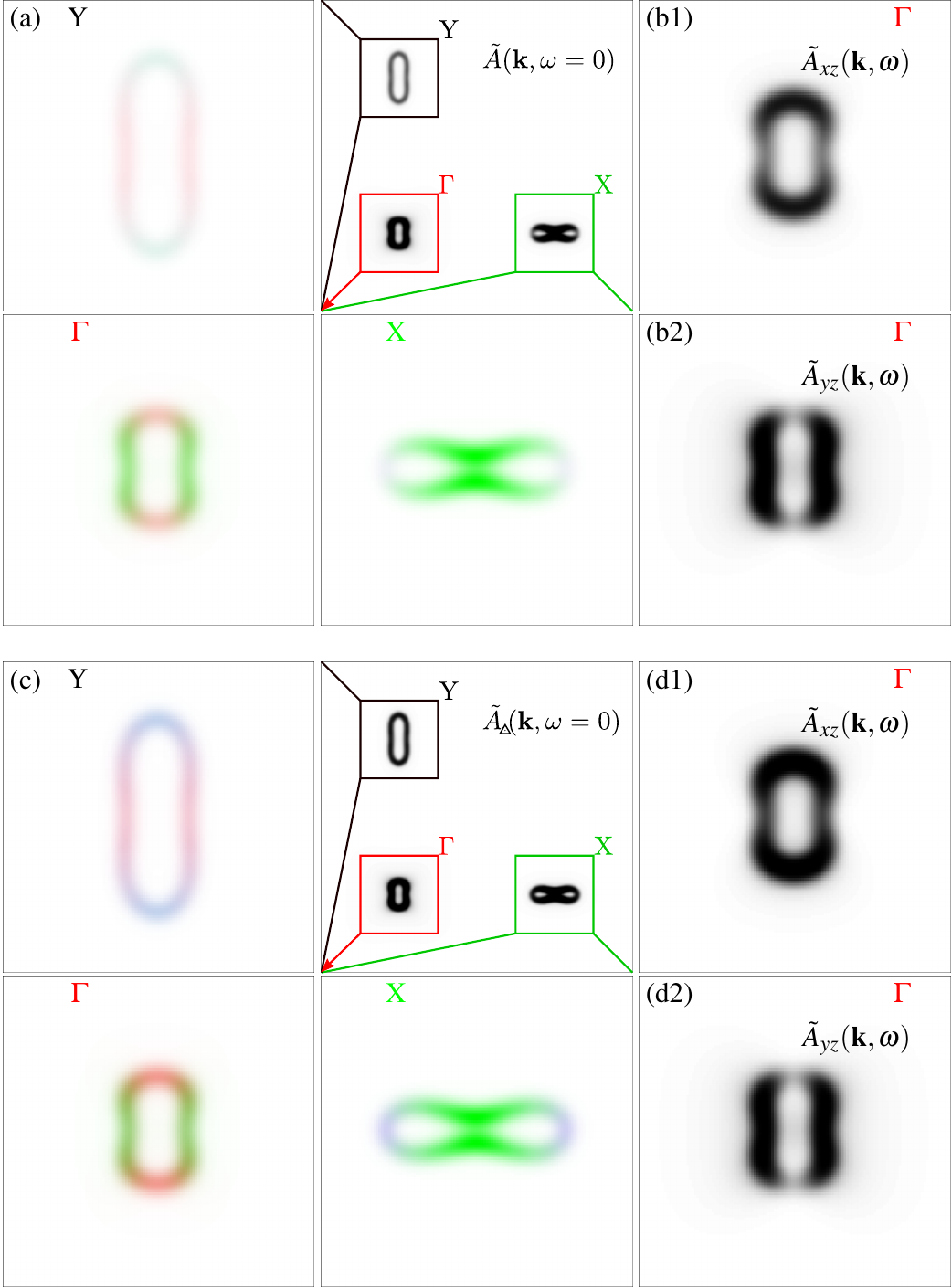}
\caption{(a) Spectral function of the electronic structure at zero energy of our model with reduced coherence ($Z_l$ set I). For better visibility, the areas around the $\Gamma$, X and Y point are blown up as indicated. (b) Partial spectral function at zero energy close to the $\Gamma$ point in the Brillouin zone.
(c) and (d) The same quantities with quasiparticle weights set II still within the range of uncertainties, see Fig. \ref{Z_bars}, marked with a triangle which are still within the range where agreement to the experimentally observed superconducting order parameter can be achieved\cite{Kreisel2017}.
\label{fig_spectral_function}}
\end{figure}

\begin{figure}[tb]
\includegraphics[width=\linewidth]{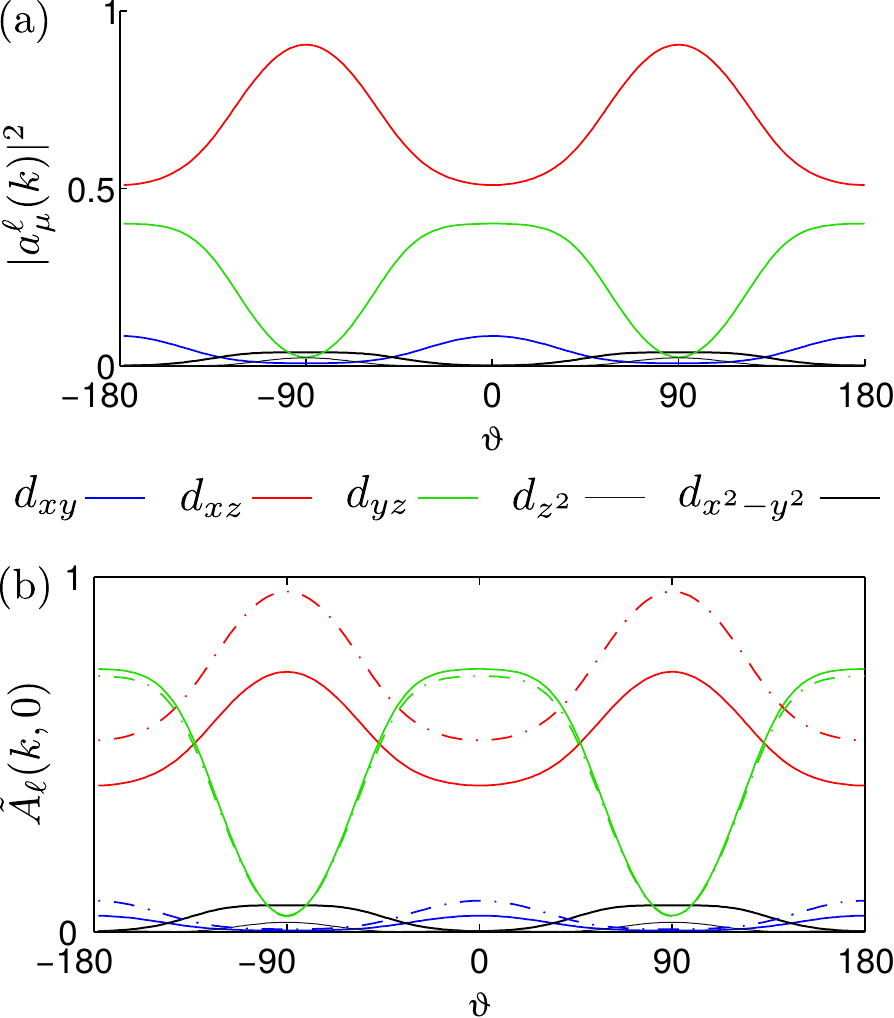}
\caption{(a) Orbital weight of the eigenfunctions $a_\mu^l(\k)$ for the points on the Fermi surface as function of the angle $\vartheta$ around the $\Gamma$ pocket. (b) Value of the orbitally resolved spectral function $\tilde A_l(\k,\omega=0)$ for the quasiparticle weights $Z_l$ set I (solid lines) and the quasiparticle weights set II with weaker correlation effects (dashed lines).
\label{fig_weights_Gamma}}
\end{figure}

\subsection{Spectral function}

The orbital-dependent quasiparticle weights $Z_l$ have direct consequences for the QPI and single-particle observables such as the spectral function, $\tilde A(\k,\omega)$. For example, in Fig.~\ref{fig_spectral_function}(a) we show $\tilde A(\k,\omega)$ at $\omega=0$ versus $\k$. As seen, the reduced quasiparticle weight of $d_{xy}$ and $d_{xz}$ compared to $d_{yz}$ leads to a washed-out $Y$-electron pocket, whereas the hole-pocket a $\Gamma$ and the peanut-shaped electron pocket at $X$ remain ``coherent''. In Fig.~\ref{fig_spectral_function}(b) we focus on the hole-pocket and compare the orbitally resolved spectral function  for $Z_\ell$ sets I and II  with $\sqrt{Z_{yz}/Z_{xz}}\approx 2.3$ (a,b) and $\sqrt{Z_{yz}/Z_{xz}}\approx 1.7$ (c,d), respectively. Both these ratios are within the regime of relative weights that fit the QPI data both in the normal state and the superconducting states, and yield a SC gap structure consistent with experiments. As seen, in the former case the hole pocket is dominated by $d_{yz}$ orbital character, whereas in the latter case $d_{xz}$ dominates the hole pocket. The reason that $d_{xz}$ may dominate despite $Z_{yz}/Z_{xz}>1$ is because of the dominant $d_{xz}$ character of the hole pocket in the bare (orbitally ordered) band as it can be seen in the angular plots of the wavefunction weights $|a_\mu^l(\k)|^2$ as presented in Fig. \ref{fig_weights_Gamma} (a). In the same figure,  the effect of the quasiparticle weights is shown by plotting the maximum of the spectral function orbitally resolved $\tilde A_l(\k,0)$ in (b).

\subsection{Spin excitations in the normal state}

\begin{figure}[b]
\includegraphics[width=\linewidth]{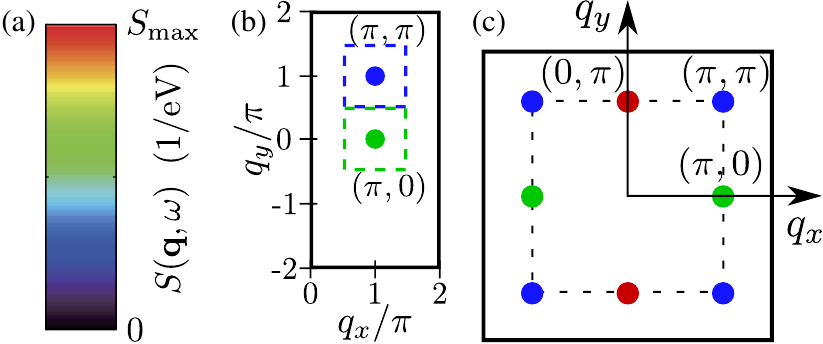}
\caption{(a) Color code for the maps of the spin fluctuations within this work. (b) Plot range of the maps and integration areas for the local fluctuations of stripe type (green box) and N\'eel type (blue box). (c) Plot range for the untwinned maps centered around the Brillouin zone (dashed line).
\label{fig_maps}}
\end{figure}

For the calculation of the spin excitations, we assume local repulsive  interactions described by the usual Hubbard-Hund Hamiltonian,\cite{Kontani2010,Graser2009,Kreisel2017} for  $U=0.57\,\text{eV}$ ($U=0.36\,\text{eV}$ for the fully coherent calculation\cite{Kreisel2017}), $J=U/6$, and calculate the static and dynamic susceptibility  $\chi_{\text{RPA}}({\bf q},\omega)$. {\colorb Note that these values of the Hubbard interaction need to be considered as effective interactions in an RPA approach and cannot be compared to the bare interactions as employed in quantum monte carlo investigations for example. These are different for the two cases in consideration, but in both cases chosen to be close to the magnetic instability as suggested by the proximity of the real FeSe system to a magnetic order.}
No additional $Z$-factors  beyond those implicitly appearing in (\ref{eqn:RPA}) enter at this level. To compare directly to the measured dynamical structure factors found by neutron scattering experiments, we use the magnetic form factor\cite{TabCrys} $f({\bf q})$ of $\text{Fe}^{2+}$ in

\begin{equation}
S({\bf q},\omega)= \frac{1}{1-e^{-\omega/T}}f^2({\bf q})\mathop{\text{Im}}\chi_{\text{RPA}}({\bf q},\omega).
\end{equation}
Note that due to the form factor, this quantity is not identical in the first and second Brillouin zones. For all the results presented below the color code shown in Fig.~\ref{fig_maps} is used and the momentum regions of the constant energy cuts are defined in the same figure.
\subsection{Twinned crystals}

\begin{figure*}[tb]
\includegraphics[width=\linewidth]{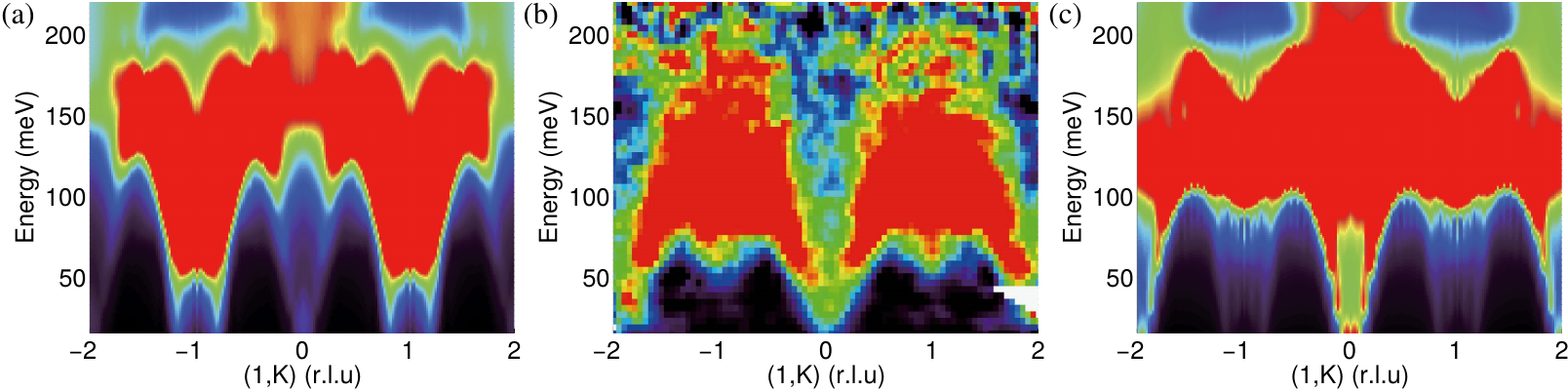}
\caption{Intensity plot of $S({\bf q},\omega)$ for a twinned FeSe crystal  showing the dispersions of the spin excitations (a) from the model without correlations,  $Z_\ell=1$ (b) measured dispersions from Ref.\onlinecite{Wang2015GS}, and (c) calculated from the model with orbital selective correlations, $Z_\ell<1$  (set I). For color scale, see Fig. \ref{fig_maps}.  Theoretical spectra were calculated at $T=40\,\text{K}$.
\label{fig_dispersion_spin_fl}}
\end{figure*}

Focusing first on the dispersion of the spin excitations relevant for twinned systems along the high symmetry directions of the Brillouin zone,  for a direct comparison with Ref. \onlinecite{Wang2015GS}, we show in Fig.~\ref{fig_dispersion_spin_fl} an intensity plot of the structure factor for the fully coherent calculation (a), the experimental result (b), and the  case   with reduced orbital coherence  ($Z_\ell$ set I) as in Ref. \onlinecite{Sprau2017} (c). As seen, the overall dispersion of the low-energy excitations is only captured in (c), where a spin- pseudogap reminiscent of the experimental situation exists near the $(\pi,\pi)$ region  and the stripe-like excitations completely dominate over the N\'{e}el-like ones.
Note that the details of the high-energy magnetic fluctuations cannot be expected to be theoretically captured by a  model  with tight-binding Hamiltonian that is known to fit mainly the low-energy band structure.    In addition, we have also not accounted for  
       the expected increase of $Z$-factors with energy.

\begin{figure*}[tb]
 \includegraphics[width=\linewidth]{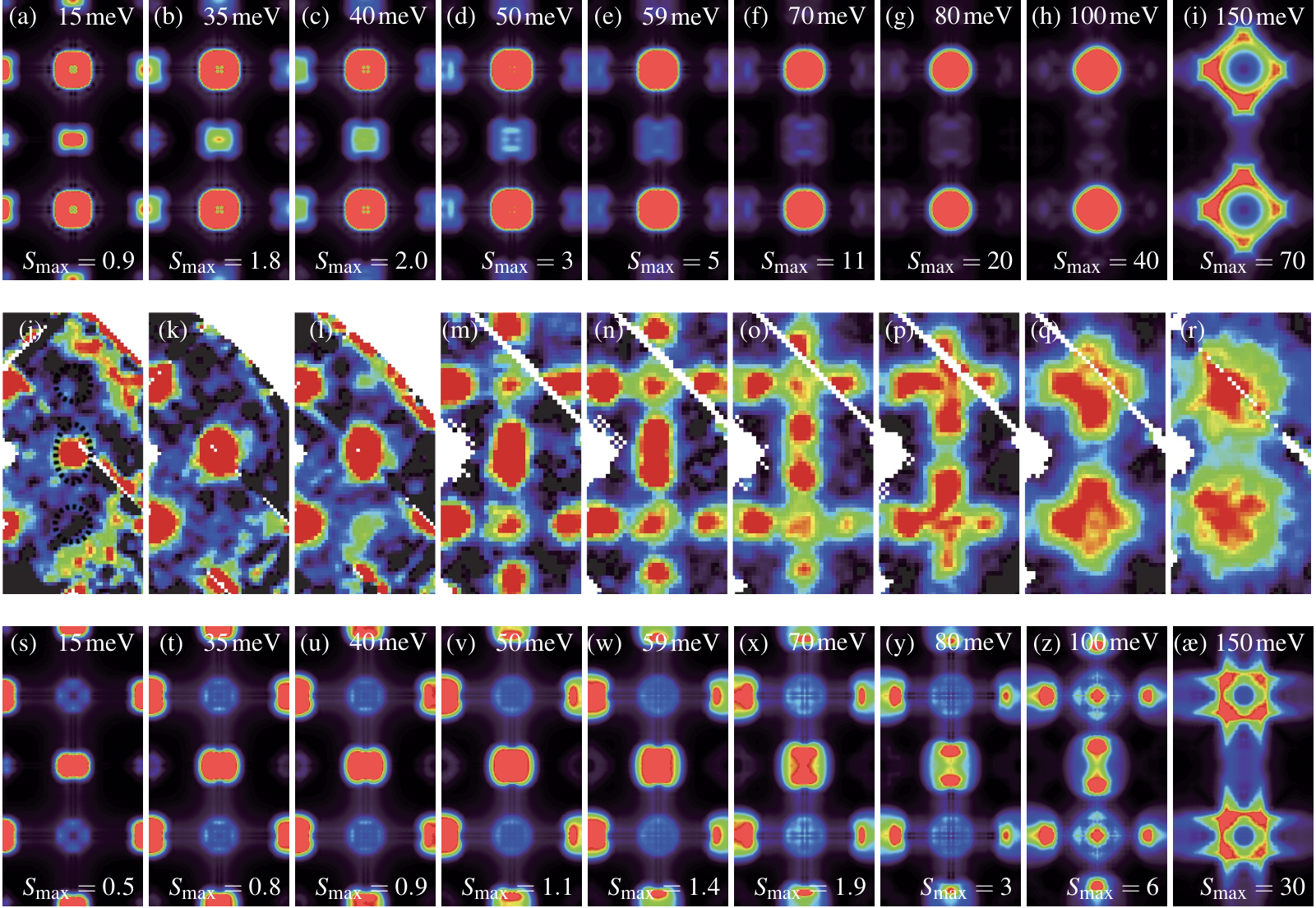}
\caption{Maps of the dynamical structure factor $S(\mathbf{q},\omega)$ at constant energies and $C_4$-symmetrized to mimic twinned crystals: (top row) The electronic structure of fully coherent electrons produce substantial spectral weight close to $(\pi,\pi)$ together with a rather weak dispersing weight close to $(\pi,0)$. The color scale is defined with the maximum set by the values given at each panel of the maps. Definition of the momentum space area and the color scale is given in Fig. \ref{fig_maps}.
(middle row) Constant energy spin fluctuations as measured by INS adapted from Ref.~\onlinecite{Wang2015GS} at the same energies. 
(bottom row) Theoretical calculation of the $C_4$-symmetrized dynamical structure factor $\tilde S(\mathbf{q},\omega)$ within the picture of orbital selective less coherent quasiparticles, the dominant intensity is exhibited close to the $(\pi,0)$ and $(0,\pi)$ regions. 
\label{fig_neutron_compare}}
\end{figure*}

Turning next to the constant energy cuts relevant for twinned crystals, we show in Fig.~\ref{fig_neutron_compare} the energy evolution of the momentum dependent spin structure factor $S(\mathbf{q},\omega)$ at low $T$. Panels (a-i) are the results for $S(\mathbf{q},\omega)$ of an uncorrelated calculation with all $Z=1$ and should be compared to panels (j-r) which show the equivalent experimental data adapted form Ref.~\onlinecite{Wang2015GS}. As seen, despite the fact that the model used to obtain the results in (a-i) exhibits a Fermi surface that almost quantitatively agrees with that extracted from ARPES and STM measurements, the agreement is poor. The model strongly overestimates the spectral weight at $\q=(\pi,\pi)$, particularly prominent at the lowest energies. The $\q=(\pi,\pi)$ weight is an inevitable consequence of the two electron pockets and their inter-band nesting of mainly $d_{xy}$ character. 

Turning next to the scenario of orbital selectively, we show in the lower row of panels in Fig.~\ref{fig_neutron_compare}(s-\ae), the theoretical result of the $C_4$-symmetrized dynamical structure factor $\tilde S(\mathbf{q},\omega)$ including the weight factors set I. 
As seen, the suppression of the $d_{xy}$ orbital contribution immediately resolves the ``$(\pi,\pi)$-problem'' at low energies. We stress that it is indeed the $d_{xy}$ orbital that exhibits the smallest $Z$-factor as found by DFT+DMFT and slave-spin methods\cite{Yi2015,Bascones_review}. The evolution of the low-energy magnetic excitations within the orbital selective scenario is in overall good agreement with the neutron data as seen from a comparison of panels (j-r) and (s-\ae). In particular one notices the dispersion of both the $\q=(\pi,0)$ and $\q=(0,\pi)$ peaks towards the $\q=(\pi,\pi)$ point around 150 meV. 

\begin{figure*}[tb]
 \includegraphics[width=\linewidth]{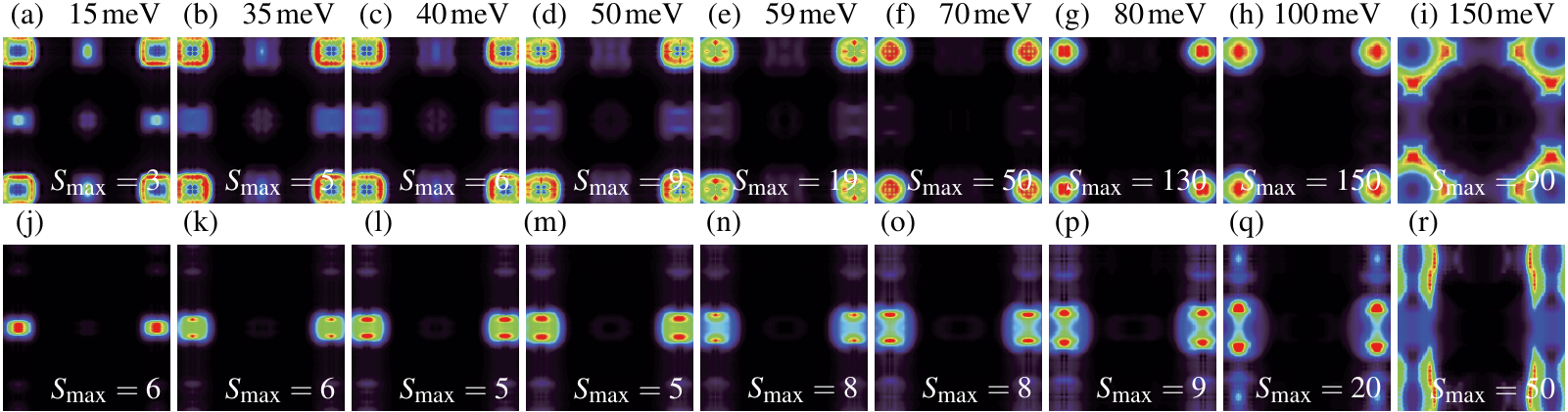}
\caption{Predicted maps of the dynamical structure factor $S(\mathbf{q},\omega)$ for inelastic neutron experiments using {\bf{untwinned}} crystals: Within the picture of fully coherent (a-i) and not fully coherent (j-r) quasiparticles presented for energy transfers as indicated in the first row and for a momentum space with $\q=0$ at the center of each map. As seen, the main difference between the two approaches is the strongly reduced weight at both $(\pi,\pi)$ and $(0,\pi)$ for the correlated case in panels (j-q). The definition of the momentum space area and the color scale is given in Fig. \ref{fig_maps}.
\label{fig_neutron_untwinned}}
\end{figure*}

\begin{figure}[tb]
\includegraphics[width=\linewidth]{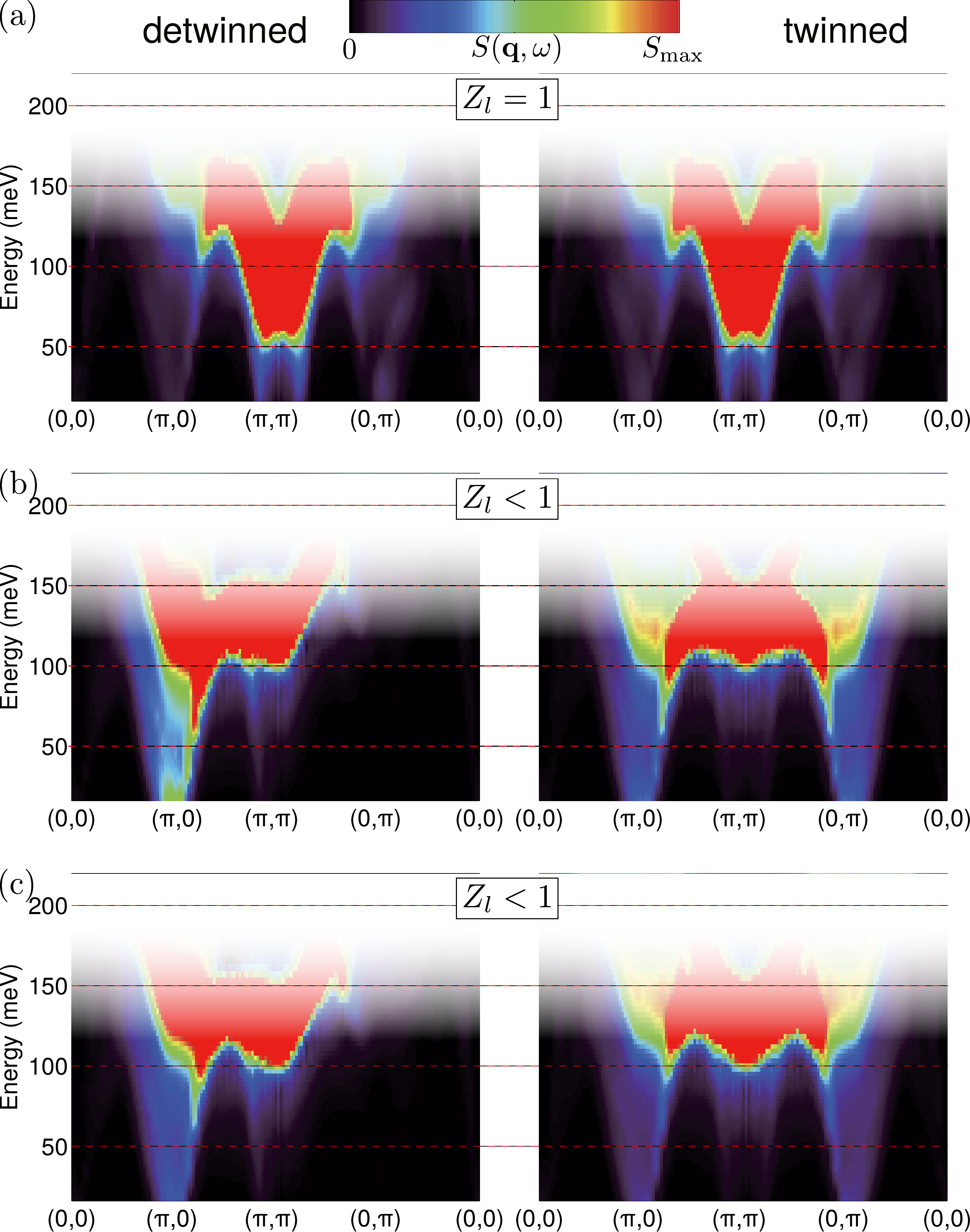}
\caption{Expected cuts in q-space of the dynamical structure factor $S(\mathbf{q},\omega)$. (left) detwinned crystals, (right) twinned crystals where $(\pi,0)$ and $(0,\pi)$ become equivalent because of the domain averaging. At higher energies, our theoretical approach is not valid any more, see text (fading out towards white). Uncorrelated model (a) compared to correlated model (b) and a calculation with an alternative set of quasiparticle weights $Z_l$ (c).
\label{fig_neutron_q_cut}}
\end{figure}

 \subsection{Untwinned crystals}
 
There are important effects of the orbital dependent quasiparticle weights $Z_l$ manifesting themselves clearly in the untwinned (non-$C_4$-symmetrized) case,  related to the surprisingly large $Z_{xz}-Z_{yz}$ splitting reported by Sprau \textit{et al}.\cite{Sprau2017} in the nematic phase. Figure~\ref{fig_neutron_untwinned} shows  constant energy cuts of $\tilde S(\mathbf{q},\omega)$ as in Fig.~\ref{fig_neutron_compare}, but not explicitly $C_4$-symmetrized  and plotted over a different $\bf q$-range to make the asymmetry  between $(\pi,0)$ and $(0,\pi)$ apparent. This models the expected neutron response at low $T$ from untwinned FeSe samples. As seen from Fig.~\ref{fig_neutron_untwinned}(j-q) the response  strongly breaks $C_4$ symmetry, such that the entire low-energy spectral weight is located at $\q=(\pi,0)$, and not at $\q=(0,\pi)$. This is in contrast to the result for a fully coherent Fermi liquid with $Z_l=1$ as seen in Fig.~\ref{fig_neutron_untwinned}(a-i).  Naturally, in both cases the structure factor is only $C_2$-symmetric due to the nematic order, but the complete absence of weight at $\q=(0,\pi)$ is not obtained from a standard itinerant scenario with electron pockets both at $X$ and $Y$. Only when including the effects of reduction of coherent weight on the $Y$-pocket, does the $\q=(0,\pi)$-signal essentially vanish. 

The dispersion of the magnetic excitations comparing the results for the detwinned and twinned cases are shown in Fig.~\ref{fig_neutron_q_cut}. Figure~\ref{fig_neutron_q_cut}(a) shows the case of fully coherent quasiparticles, whereas Fig.~\ref{fig_neutron_q_cut}(b) displays the case of orbital selective quasiparticle weights and (c) the same result for the alternative set of $Z_l$.  Only in the latter two cases, is the strong anisotropy clearly evident for the detwinned situation. The final excitation spectrum is essentially gapped except from the branch dispersing down towards $\q=(\pi,0)$. {\colorb Note that our approach should only be reliable in the low energy regime which we made evident in Fig.~\ref{fig_neutron_q_cut} by fading out the colors above $\sim 100\,\text{meV}$. Beyond this energy, we do not have a reliable model for the electronic structure (eigenenergies  $\tilde E_\mu({\bf k})$), the incoherent part of the Green's function might be non-negligible and the broadening of the coherent part of the Green's function due to an imaginary part of the self-energy cannot be ignored any more.}

\begin{figure}[tb]
  \includegraphics[width=\linewidth]{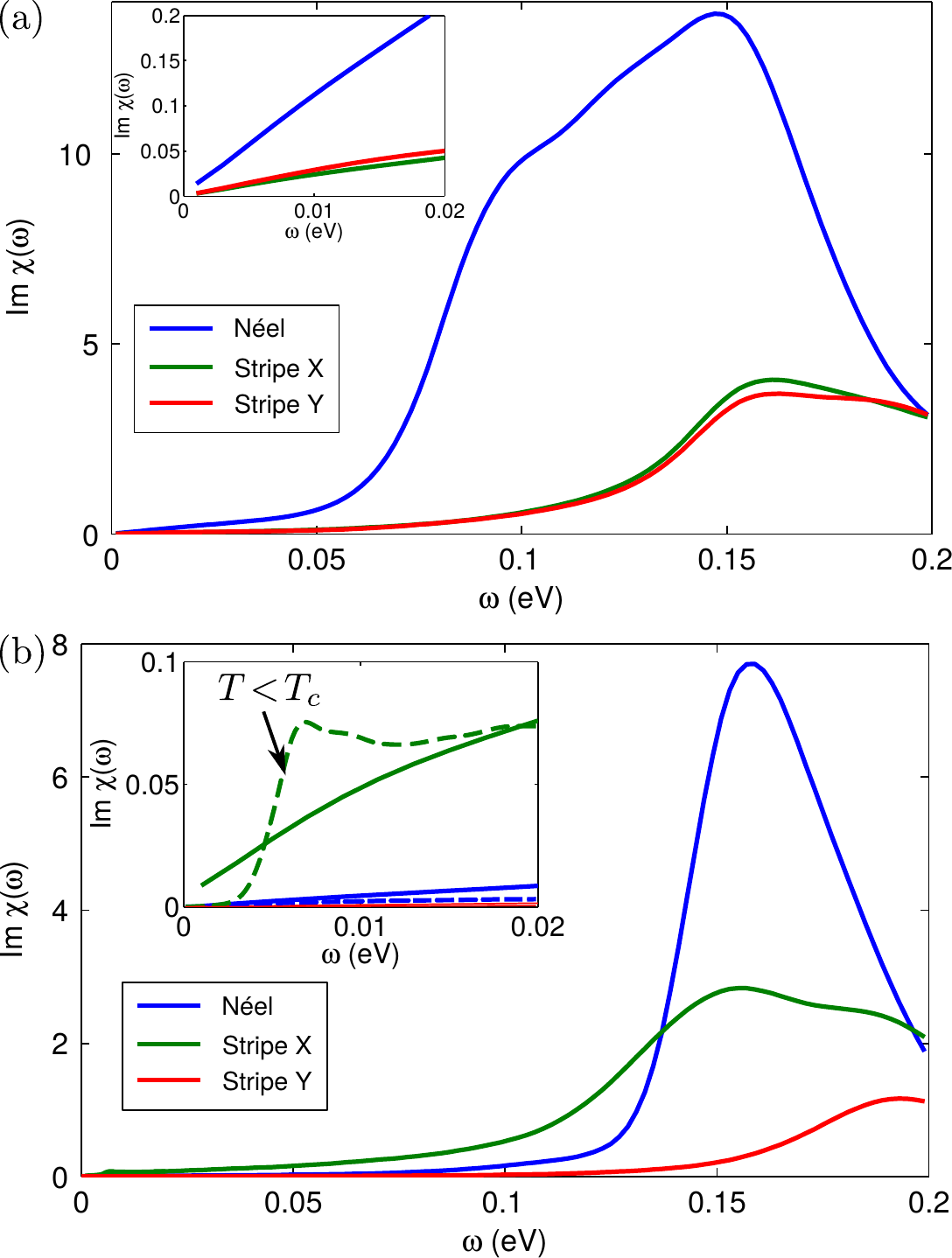}
\caption{Momentum integrated spin fluctuations separating fluctuations at $(\pi,0)$ (stripe X), $(0,\pi)$ (stripe Y) and $(\pi,\pi)$ (N\'eel) for the fully coherent model (a) and the corresponding result with reduced coherence (b). The insets show details at low energy.
\label{fig_fluct}}
\end{figure}

The shift in the momentum structure of the spectral weight between the coherent and less-coherent cases can be quantified in momentum integrated spin fluctuations separating fluctuations at $(\pi,0)$ (stripe X), $(0,\pi)$ (stripe Y) and $(\pi,\pi)$ (N\'eel), and is displayed in Fig.~\ref{fig_fluct}. As seen, the stripe X and stripe Y regions are essentially identical (vastly distinct) in Fig.~\ref{fig_fluct}(a) (Fig.~\ref{fig_fluct}(b)) throughout the whole low energy region. The inclusion of quasiparticle weights imply that the spin fluctuations at the lowest energies are strongly dominated by the $(\pi,0)$ region as seen by comparison of Fig.~\ref{fig_fluct}(a) and \ref{fig_fluct}(b). 
 \begin{figure}[tb]
  \includegraphics[width=\linewidth]{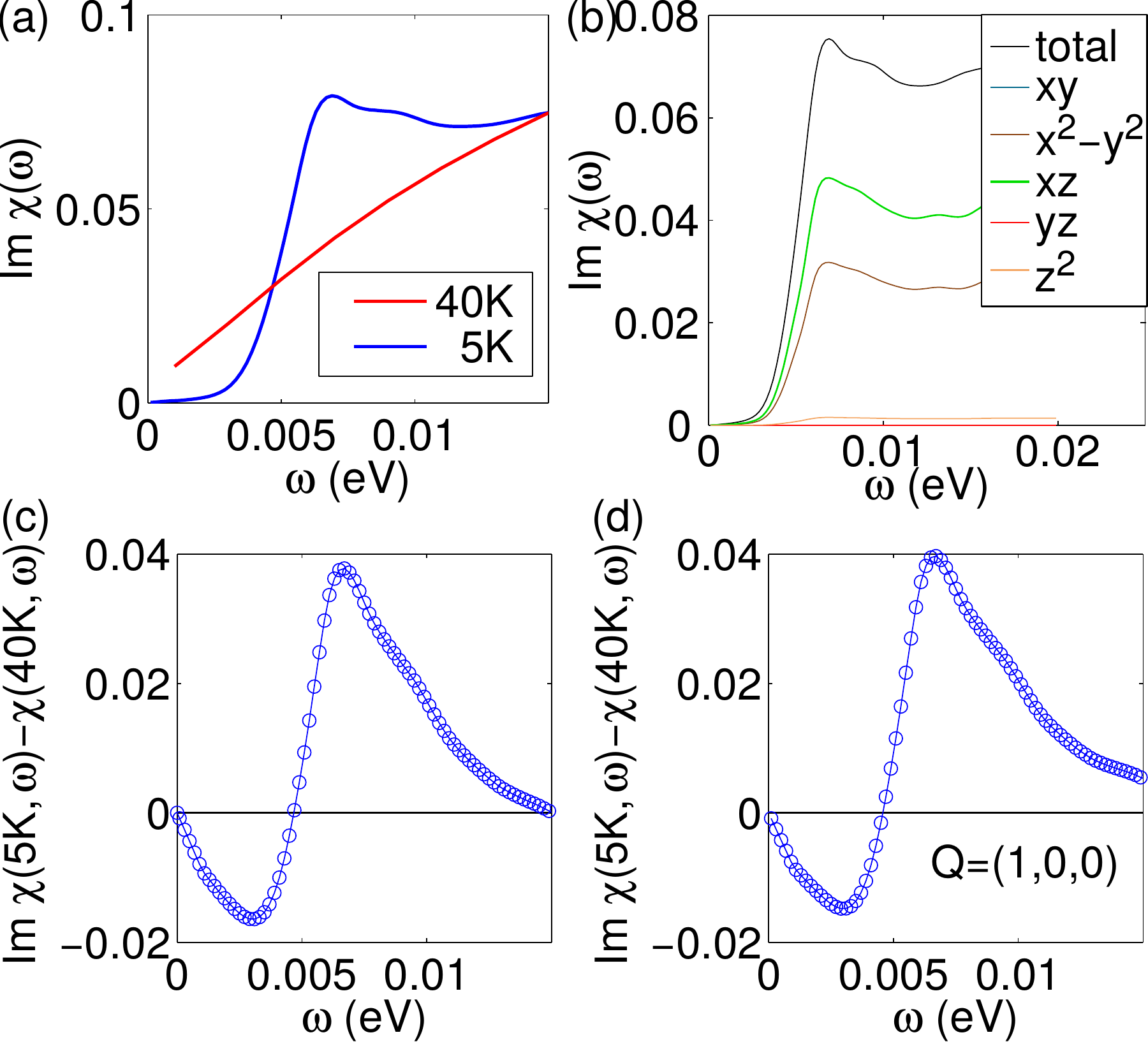}
\caption{Plot of the magnetic susceptibility to show the effect of the superconducting state: (a) Local susceptibility (integrated over the full BZ) at high temperature and low temperature, (b) orbitally resolved local susceptibility at low temperature, (c) difference between $T=40\,\text{K}$ simulation and $T=5\,\text{K}$ simulation; (d) same as before, but only integrated around $(\pi,0)$, the dominating momentum transfer.
\label{fig_fluct_relative}}
\end{figure}

\subsection{Response in the superconducting state}
In the superconducting state, we calculate the BCS-susceptibility tensor by assuming an order parameter that is diagonal in band space. For this purpose, we use the result of a calculation of the symmetry function $g(\k)$ on the Fermi surface from the solution of the linearized gap equation\cite{Sprau2017,Kreisel2017} and approximate the order parameter away from the Fermi surface using a Gaussian damping factor of $\exp(-\tilde E_\mu(\k)^ 2/\Delta E^2)$ where the energy scale $\Delta E$ is several times the maximum gap magnitude\cite{Maier09}. Then, we can obtain the eigenenergies of the Bogoliubov quasiparticles by separate diagonalizations in each band to get $\epsilon_\mu(\k)=\sqrt{\tilde E_\mu(\k)^2+\Delta_\mu(\k)^ 2}$. The corresponding expression for the susceptibility tensor is then\cite{Maier09}

\begin{align}
 &\tilde\chi_{\ell_1 \ell_2 \ell_3 \ell_4}^0 (q)  = - \sum_{k,\mu,\nu }\tilde M_{\ell_1 \ell_2 \ell_3 \ell_4}^{\mu\nu} (\k,\q)\\
 &\times[ G^{\mu}_{\text{BCS}} (k+q)  G^{\nu}_{\text{BCS}} (k)+ F^{\mu}_{\text{BCS}} (k+q)  F^{\nu}_{\text{BCS}} (-k)], \notag  \label{eqn_BCS_tensor}
\end{align}
where the normal Green's function reads 
\begin{equation}
G^{\mu}_{\text{BCS}} (k)=\frac{i\omega_n+\tilde E_\mu(\k)}{\omega_n^2+\epsilon_\mu(\k)^2},
\end{equation}
and the anomalous Green's function is given by
\begin{equation}
 F^{\mu}_{\text{BCS}} (k)=\frac{\Delta_\mu(\k)}{\omega_n^2+\epsilon_\mu(\k)^2}\,.
\end{equation}
In Fig.~\ref{fig_fluct_relative}, we focus on the low-energy region and the emergence of the neutron resonance as seen by the blue curve in Fig.~\ref{fig_fluct_relative}(a).  The neutron resonance occurs only at $(\pi,0)$,
as seen both from the inset in Fig.~\ref{fig_fluct}(b) and comparison of the two difference-plots in Fig.~\ref{fig_fluct_relative}(c,d). An unusual property of the FeSe neutron resonance within the present scenario is seen from the orbital resolved spin susceptibility in Fig.~\ref{fig_fluct_relative} (b), i.e. it is mainly of $d_{yz}$ character. Therefore, an important consequence of the orbital selective quasiparticle weights is to render the neutron resonance highly momentum and orbital dependent.

\section{Relation to other works}
Iron selenide is a fascinating system where nematicity, magnetic fluctuations,  and significant electronic correlations conspire to produce unusual superconductivity in a compensated metal with tiny Fermi surface pockets of varying orbital character.  The small energy scales and orbital mixing make the compound a significant challenge for both theory and experiment.  Since the discovery of new cold vapor deposition methods allowed for the growth of high-quality stoichiometric crystals several years ago, theories have been gradually improved as higher resolution experiments were performed.  Recently, several theories\cite{Kontani_PRL_FeSe,She2017arXiv,Kang2018,Eugenio2018,Benfatto2018,Rhodes2018} in addition to the present approach\cite{Sprau2017,Kreisel2017} have appeared which study self-energy effects in FeSe, and some also present calculations for the structure of the superconducting gap measured by QPI and  ARPES\cite{Liu2018,Rhodes2018,Hashimoto2018,Kushnirenko2018}.  In many respects, these ideas parallel our own, although they do not all invoke the concept of quasiparticle weight renormalization.  The similarity arises because within our approximation, the quasiparticle weight factors $Z_\ell$ influence the effective interaction  by renormalizing its momentum dependence.  In this framework, the incoherent part of the electron is not included, an assumption that needs to be justified by more complete microscopic calculation, but seems reasonable provided the incoherent weight is at high energies\cite{Watson_hubbard2017,Evtushinsky2016}.

Recent alternative theoretical works have discussed other possible causes for an unusual momentum dependence of the effective interaction that ultimately gives similar results for the superconducting gap.   Kang \textit{et al}.\cite{Kang2018} considered a low-energy model\cite{Cvetkovic2013} with independent orbital ordering amplitudes at $\Gamma/Z$, $X$ and $Y$ points\footnote{Eugenio and Vafek\cite{Eugenio2018} have recently used the same expansion of the electronic structure around high symmetry points, and classified all possible pairing states within the multiorbital mean field theory.}, and calculated the effect of this induced nematicity on the quasiparticle orbital character at the Fermi surface. They did not include quasiparticle decoherence effects, with the exception of the $d_{xy}$ orbital, which they completely suppressed explicitly.  These authors then found electron and hole weights in accord with some ARPES experiments, and deduced anisotropic repulsive interactions leading to a superconducting gap with the correct qualitative anisotropy if calculated using an electronic structure with large hole pocket as observed close to the $Z$ point. The anisotropy of the gap is however very small and has minima where maxima are observed experimentally if the gap is calculated on the small hole pocket near the $\Gamma$ point. Benfatto \textit{et al.}\cite{Benfatto2018} followed an approach which has similar physical consequences as the one presented in the present work. The main difference is that the deformation of the low-energy electronic structure in the nematic phase is described by calculating a one-loop self-energy based on a phenomenological nematic spin fluctuation propagator. The corresponding pairing interaction was found to be highly anisotropic, leading to a gap structure consistent with experiment.  
Rhodes \textit{et al.}\cite{Rhodes2018}  accomplished the same thing in a nominally more realistic RPA spin fluctuation pairing calculation by ignoring the contribution from the Fermi surface pocket at $Y$ to the pairing entirely, while nevertheless including this band in the quasiparticle energy.    While apparently inconsistent, this amounts to a theory of strong pocket- rather than orbital-dependent quasiparticle renormalizations somewhat analogous to the approach presented here and in Refs. \onlinecite{Sprau2017,Kreisel2017}.

All these itinerant approaches propose fits to the gap anisotropy, but since full calculations of the dynamical susceptibility are not always presented,  it is not clear what predictions they would make for the relative intensities of $(\pi,0)$ and $(0,\pi)$ excitations.  We therefore propose this as a good test to distinguish between the various theoretical models. We note that the strong-coupling theory of the ``nematic quantum paramagnet'' put forward in Ref. \onlinecite{She2017arXiv} reaches similar conclusions to ours regarding the $(\pi,0)/(0,\pi)$ anisotropy. 
Some degree of anisotropy would generally also be obtained within the framework of  Ref. \onlinecite{Benfatto2018}, which makes the basic assumption of stronger fluctuations at $(\pi,0)$.

Since the publication of Ref. \onlinecite{Sprau2017}, new experiments have both confirmed and challenged some aspects of our description.  The quasiparticle interference experiment of Ref. \onlinecite{Kostin2018} shows that the momentum dependent renormalization of interactions is present already in the nematic normal state above $T_c$, and is remarkably consistent with the $Z_\ell$-factors determined by fit to experiment on the superconducting gap structure.  The $Y$ electron pocket, which should have little coherent spectral weight in our analysis since it consists entirely of $d_{xy}$ and $d_{xz}$ weight, has proven nearly impossible to observe in either ARPES\cite{Watson2017,Rhodes2018} or STM\cite{Sprau2017}.
     
On the other hand several ARPES experiments have found less $d_{yz}$ weight on the $\Gamma$-centered hole pockets than expected in the current renormalized band structure\cite{Liu2018,Rhodes2018,Watson2017}, and failed to detect the expected reduction in $d_{xy}$ and $d_{xz}$ character on electron pockets\cite{Rhodes2018,Kushnirenko2018}.  We believe that the actual determination of the $Z$ factors from ARPES measurements will require careful modeling and  integration of Energy Distribution Curves (EDCs), and will be needed to resolve these controversies. 

\section{Conclusions}
  Above, we reviewed a theory of the low-energy bandstructure and interactions that accounts only for the coherent part of the quasiparticles near the Fermi level, with strongly reduced $xy$ and $xz$ quasiparticle weights, which earlier had been shown to account well for the structure of the superconducting gap and normal state quasiparticle interference. 
In this work, we have calculated the expected inelastic neutron scattering intensity for FeSe within this framework, considering both the set of $Z_\ell$'s taken from Ref. \onlinecite{Sprau2017}, as well as a set corresponding to more moderate correlations.  Both are consistent with the low energy, low-temperature neutron data, as well as with the superconducting gap anisotropy within experimental error bars.  As expected, the strong decoherence of the $d_{xy}$ states at the Fermi level dramatically suppresses 
$(\pi,\pi)$ excitations at low energy, leading in fact to a spin pseudogap roughly consistent with experiment.  In addition, we have made predictions for untwinned crystals appropriate for when such experiments become feasible.    The very striking aspect of the theory is that the     low-energy spin excitation intensity at $(\pi,0)$ and $(0,\pi)$ becomes extremely anisotropic, with $(0,\pi)$ fluctuations essentially eliminated.     

That this is {\it not} the case for the standard RPA approach without such quasiparticle renormalizations was pointed out earlier in Ref. \onlinecite{She2017arXiv}, which obtained similar results to the present work regarding this point within a strong-coupling picture.  This agreement suggests that our renormalized multiorbital Fermi liquid shares features in common with the nematic quantum paramagnetic state described in that work, which was inspired in turn by explanations of the lack of long-range magnetic order in FeSe relying on frustration of various competing  magnetic states\cite{Glasbrenner2015,WangLee2015NatPhys_FeSe-Para-Nematicity}. {\colorb It is however difficult to make a direct connection between our Fermi liquid picture and the quantum paramagnetic state because the latter does not adopt a perturbative approach.}

In addition, we have discussed the results of several other itinerant approaches to FeSe, all of which have proposed slightly differing accounts of how the observed gap anisotropy arises.  That this is possible follows from the fact that our approach is essentially equivalent to a $\k$-dependent renormalization of the pairing interaction, in particular one which suppresses $(\pi,\pi)$ and $(0,\pi)$ fluctuations and thereby the pairing amplitudes in these directions.  Other approaches can in principle construct effective interactions for fully coherent quasiparticles where the required momentum dependence arises from a redistribution of orbital character in the nematic phase, or from highly anisotropic nematic spin fluctuations.   At present, measurements of the superconducting properties, given apparent disagreements among ARPES measurements, are probably not sufficient to enable one to differentiate among the theories.  We have therefore proposed here that inelastic neutron scattering measurements on untwinned FeSe  will be a fundamental and useful experimental probe addressing this question directly.  The orbitally selective spin fluctuation approach, with values of quasiparticle weights $Z_\ell$ determined by fits to both superconducting state\cite{Sprau2017} and normal state QPI\cite{Kostin2018},  implies a very large $(\pi,0)/(0,\pi)$     anisotropy in the spin excitation spectrum of such a system.  This prediction should be observable in inelastic neutron measurements on mechanically detwinned samples.

\section{Acknowledgments}

We acknowledge useful discussions with  S. Backes, L. Benfatto, A. E. B\"ohmer, T. Chen, A. Chubukov, P. Dai, J.C. Davis, L. Fanfarillo, R. Fernandes, J. Kang, A. Kostin, D. D. Scherer, and P.O. Sprau . B. M. A. acknowledges support from the Independent Research Fund Denmark grant number DFF-6108-00096. P. J. H.  was supported by the Department of Energy under Grant No. DE-FG02-05ER46236.


%

\end{document}